\documentclass[aps,prl,reprint,groupedaddress,floatfix]{revtex4-2}

\usepackage[T1]{fontenc}
\usepackage{newtxtext}
\usepackage{newtxmath}

\usepackage{amsmath}
\usepackage{mathtools}
\usepackage{amsfonts}

\usepackage{tikz}
\usetikzlibrary{quantikz}

\usepackage{color}
\usepackage{xcolor}
\usepackage[colorlinks, citecolor=blue, linkcolor=blue, urlcolor=blue]{hyperref}

\usepackage{graphicx}
\usepackage[all]{hypcap}

\begin{document}

\title{Quantum criticality using a superconducting quantum processor}

\author{Maxime Dupont}
\affiliation{Department of Physics, University of California, Berkeley, California 94720, USA}
\affiliation{Materials Sciences Division, Lawrence Berkeley National Laboratory, Berkeley, California 94720, USA}

\author{Joel E. Moore}
\affiliation{Department of Physics, University of California, Berkeley, California 94720, USA}
\affiliation{Materials Sciences Division, Lawrence Berkeley National Laboratory, Berkeley, California 94720, USA}

\begin{abstract}
    Quantum criticality emerges from the collective behavior of many interacting quantum particles, often at the transition between different phases of matter. It is one of the cornerstones of condensed matter physics, which we access on noisy intermediate-scale (NISQ) quantum devices by leveraging a dynamically driven phenomenon. We probe the critical properties of the one-dimensional quantum Ising model on a programmable superconducting quantum chip via a Kibble-Zurek process, obtain scaling laws, and estimate critical exponents despite inherent sources of errors on the hardware. In addition, we investigate how the improvement of NISQ computers (more qubits, less noise) will consolidate the computation of those universal physical properties. A one-parameter noise model captures the effect of imperfections and reproduces the experimental data. Its systematic study reveals that the noise, analogously to temperature, induces a new length scale in the system. We introduce and successfully verify modified scaling laws, directly accounting for the noise without any prior knowledge. It makes data analyses for extracting physical properties transparent to noise. By understanding how imperfect quantum hardware modifies the genuine properties of quantum states of matter, we enhance the power of NISQ processors considerably for addressing quantum criticality and potentially other phenomena and algorithms.
\end{abstract}

\maketitle

The advent of quantum computing promises to disrupt nearly every industry, from materials science, chemistry, and drug discovery to security, optimization, as well as artificial intelligence. However, current quantum processors have limited computing capabilities, with only a small number of imperfect qubits available. Although quantum advantage~\cite{Harrow2017} has been claimed on such NISQ devices~\cite{Arute2019,Zhong2020}, it is only on specific tasks of narrow interest. Therefore, a major goal is to address practical problems with NISQ machines~\cite{Preskill2018}. Quantum many-body problems, which seek to describe interacting quantum degrees of freedom, provide an ideal playground. Not only are they suitable for current and future NISQ hardware, but they are also of prime importance in basic research. They span nuclear, high-energy, condensed matter, atomic, molecular, optical physics, and quantum chemistry. Only a corner of quantum many-body problems can be solved efficiently with classical computers---these can serve for benchmarking---whereas the vast majority is still open.

For instance, competing interactions between quantum particles can lead to the emergence of exotic phases of matter and phase transitions between them~\cite{Vojta2003,Sachdev2011}. Of particular interest are quantum many-body systems experiencing a second-order quantum phase transition, as they exhibit quantum criticality~\cite{PhysRevB.14.1165,Laughlin2001,Coleman2005}: an emerging scale-invariance dictating how physical quantities (e.g., susceptibility, specific heat, spectral gap, correlations, etc.) behave close to the transition. Quantum criticality is tabulated into universality classes, defined by a set of critical exponents characterizing the nature of the transition. Remarkably, universality classes are independent of most of the microscopic details of a quantum system and depend instead on general attributes such as dimension and symmetries. Hence, accessing, classifying, and understanding quantum criticality is a formidable fundamental physics challenge. A conventional way for investigating quantum criticality in a quantum many-body system consists of studying its ground state properties as a function of a parameter $g$ driving the transition, with the transition taking place at $g=g_\mathrm{c}$, known as the quantum critical point (QCP)~\cite{Sachdev2011}. However, obtaining the lowest-energy state of a given Hamiltonian $\hat{\mathcal{H}}(g)$ is a cumbersome task for NISQ devices.

We bypass this obstacle by leveraging a dynamically driven phenomenon to access quantum criticality, the Kibble-Zurek (KZ) mechanism~\cite{Kibble1976,Zurek1985}. NISQ processors have proven to be well-suited in simulating quantum dynamics~\cite{Barends2015,Zhukov2018,PhysRevA.98.032331,PhysRevLett.121.170501,CerveraLierta2018,Chiesa2019,Smith2019,PhysRevB.101.014411,PRXQuantum.2.010342,Mi2021,Guo2021,Fauseweh2021,Geller2021,Vovrosh2021,randall2021,dumitrescu2021,mi2021b,Xu2021,zhao2021,Bassman2022,Urbanek2021,Kokcu2022}, as the time evolution is a unitary operation that can be straightforwardly translated into a shallow quantum circuit in most cases. The KZ mechanism is triggered by time evolving a system from a point A to a point B of its phase diagram at a given rate $\sim T^{-1}$, with the transition happening somewhere on the way. With the spectral gap of a quantum system vanishing as $\Delta\sim|g-g_\mathrm{c}|^{z\nu}$ close to a second-order phase transition (with $z,~\nu>0$ the dynamical and correlation length critical exponents, respectively)~\cite{Vojta2003,Sachdev2011}, we expect a characteristic timescale $\tau$ and associated gap scale $\hbar/\tau$, where the adiabaticity of the evolution breaks down. It happens at a dimensionless distance $|g-g_c|/g_c\sim\tau/T$ of the critical point, and one finds that $\tau\sim T^{z\nu/(1+z\nu)}$. Likewise, a characteristic length scale $\ell\sim\tau^{1/z}$ emerges. It diverges in the adiabatic limit, leading to scale invariance, as one would expect in  the ground state of $\hat{\mathcal{H}}(g=g_\mathrm{c})$.

Because the KZ mechanism is controlled by the same critical exponents as the static physics, one can exploit it to access key properties of quantum criticality in many-body systems. For example, the KZ process was recently used in a Rydberg atomic simulator to study a quantum critical point~\cite{Ebadi2021}. Here, we analyze a classic example of quantum criticality in one spatial dimension through both a gate-based quantum processor and a classical matrix product state computation incorporating noise. We find that the effect of noise is analogous to that of temperature: It induces a length scale that can be accounted for through modified scaling laws. Our results enhance the power of NISQ processors significantly by making data analyses transparent to inherent noise. Understanding how imperfect quantum hardware modifies the genuine properties of quantum states of matter is a prerequisite for condensed matter simulations, which are doomed to be noisy in the near future.

\begin{figure}[t]
    \includegraphics[width=0.8\columnwidth]{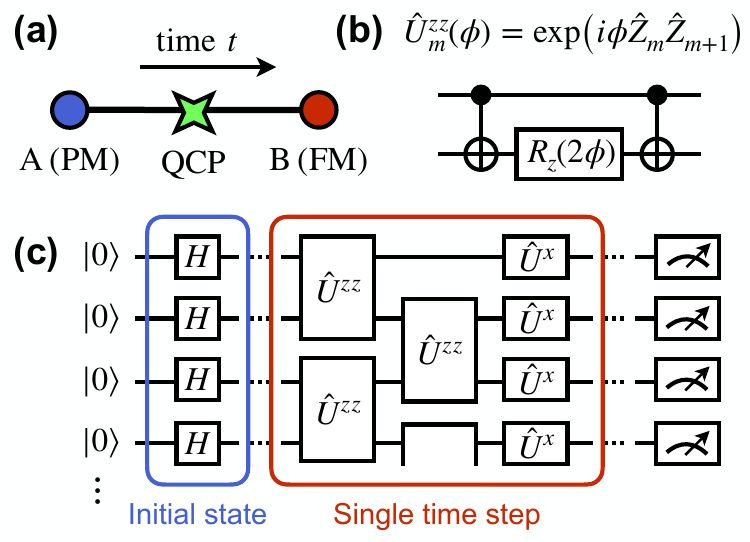}
    \caption{(a) Quantum system dynamically driven from a point A (paramagnetic phase---PM, for the quantum Ising model considered here) to a point B (ferromagnetic phase---FM) of its phase diagram, with a transition happening on the way, characterized by a quantum critical point (QCP). (b) Decomposition of the operator $\hat{U}^{zz}_m(\phi)=\exp\bigl(i\phi\hat{Z}_{m}\hat{Z}_{m+1}\bigr)$ by sandwiching a single-qubit rotation gate around the $z$ axis by two-qubit \textsf{CNOT} gates. (c) Quantum circuit for the discretized unitary operation of Eq.~\eqref{eq:time_evolution} for the quantum Ising model~\eqref{eq:ising_terms}. The PM ground state is constructed by applying Hadamard gates $H$ on individual qubits. The second step is the time-evolution, generated by a first-order Suzuki-Trotter expansion. Here, $\hat{U}^x_m(\phi)=\exp\bigl(i\phi\hat{X}_m\bigr)$.}
    \label{fig:introduction}
\end{figure}

One may write a Hamiltonian interpolating between points A and B in a KZ process as,
\begin{equation}
    \hat{\mathcal{H}}\bigl(T,t\bigr)=\bigl(1-t\bigr/T\bigr)\hat{\mathcal{H}}_\mathrm{A}+\bigl(1+t\bigr/T\bigr)\hat{\mathcal{H}}_\mathrm{B},
    \label{eq:hamiltonian}
\end{equation}
running from time $t=-T$ to $t=+T$ with $\hat{\mathcal{H}}_\mathrm{A,B}$ describing A and B, respectively. After initially preparing the system $\vert\Psi(t=-T)\rangle$ into the ground state of $\hat{\mathcal{H}}_\mathrm{A}$, it is dynamically driven to point B,
\begin{equation}
    \bigl\vert\Psi(t)\bigr\rangle=\mathcal{T}\mathrm{exp}\left[-\frac{i}{\hbar}\int^{t}_{-T}\mathrm{d}t'\,\hat{\mathcal{H}}\bigl(T,t'\bigr)\right]\bigl\vert\Psi\bigl(-T\bigr)\bigr\rangle,
    \label{eq:time_evolution}
\end{equation}
as pictured in Fig.~\ref{fig:introduction}(a). $\mathcal{T}$ indicates a time-ordered exponential. Close to the transition, i.e., around a model-dependent value of $t$, the KZ mechanism will kick in, and $\vert\Psi(t)\rangle$ will display universal quantum critical properties. They can be extracted and studied by computing standard observables supplemented with a scaling analysis~\cite{PhysRevLett.28.1516}.

We consider the quantum Ising model in one dimension, whose microscopic Hamiltonian interpolates between paramagnetic (PM) $\equiv$A and ferromagnetic (FM) $\equiv$B phases,
\begin{equation}
    \hat{\mathcal{H}}_\mathrm{PM}=-\sum\nolimits_n\hat{X}_n,~~\mathrm{and}~~\hat{\mathcal{H}}_\mathrm{FM}=-\sum\nolimits_n\hat{Z}_n\hat{Z}_{n+1},
    \label{eq:ising_terms}
\end{equation}
with $\hat{X}_n$ and $\hat{Z}_n$ as Pauli operators acting on qubit $n$. This model presents several advantages: first, it provides the standard paradigm of a solvable QCP at the transition between the two phases. Second, its dynamics can be encoded as a quantum circuit with a relatively low gate count. Third, in the basis where $\hat{Z}$ is diagonal, the starting point (ground state of $\hat{\mathcal{H}}_\mathrm{PM}$) is an equal superposition of all basis states, which can be readily obtained by applying individual Hadamard gates on each of the qubits. With the interpolation of Eq.~\eqref{eq:hamiltonian}, it is known that the QCP is located at $t=0$. Furthermore, the KZ mechanism on the quantum Ising model is extensively documented~\cite{PhysRevLett.95.105701,PhysRevLett.95.245701,PhysRevA.75.052321,Dziarmaga2010,PhysRevA.83.062104,PhysRevA.100.032115,PhysRevLett.109.015701,Gong2016,PhysRevB.86.064304,PhysRevB.93.075134,PhysRevLett.123.130603,Keesling2019,Ebadi2021,Schmitt2021}.

The evolution operator in Eq.~\eqref{eq:time_evolution} is discretized by making the Hamiltonian operator piecewise constant over a time step $\delta t$. Thanks to the locality of the Ising terms in Eq.~\eqref{eq:ising_terms}, the exponentiation can be performed using a Suzuki-Trotter expansion~\cite{hatano2005}, at the expense of a systematic---yet controlled---error. It engenders operators of the form $\hat{U}^x_m(\phi)=\exp\bigl(i\phi\hat{X}_m\bigr)$ and $\hat{U}^{zz}_m(\phi)=\exp\bigl(i\phi\hat{Z}_{m}\hat{Z}_{m+1}\bigr)$, which can be easily translated into standard quantum logic gates. The former is directly related to a single-qubit rotation gate around the $x$ axis, $R_x(\phi)$, and the latter can be decomposed into standard gates~\cite{PhysRevA.69.032315}, see Fig.~\ref{fig:introduction}(b). The quantum circuit for one time step using a first-order Suzuki-Trotter expansion is shown in Fig.~\ref{fig:introduction}(c).

To investigate quantum criticality, we look at the two-point correlation function,
\begin{equation}
    C\bigl(T,t,x\bigr)=\bigl\langle\Psi(t)\bigr\vert\hat{Z}_r\hat{Z}_{r\pm x}\bigr\vert\Psi(t)\bigr\rangle,
    \label{eq:correlation}
\end{equation}
between a reference qubit $r$ assumed in the middle of the system and another qubit at distance $x$. Close to the QCP, it is expected to show a universal behavior of the form~\cite{PhysRevLett.109.015701,PhysRevB.86.064304,PhysRevB.93.075134},
\begin{equation}
    C\bigl(T,t,x\bigr)=\ell^{-\eta}\,\mathcal{F}\bigl(x/\ell,t/\tau\bigr),
    \label{eq:correlation_critical_scaling}
\end{equation}
with $\mathcal{F}$ being a nonuniversal scaling function, $\eta$ is the anomalous critical exponent, $\ell$ and $\tau$ the characteristic length and time scales of the KZ mechanism, which depend on $T$ and the critical exponents. $\ell$ can be interpreted as the length over which the system will be defect-free. From there, one can deduce that the adiabatic limit for a system of size $L$ will be recovered for drive times $T\gtrsim L^{(1+z\nu)/\nu}$---though the point of the KZ mechanism is that useful physics can still be extracted outside of the adiabatic regime.

\begin{figure*}[t]
    \includegraphics[width=2.1\columnwidth]{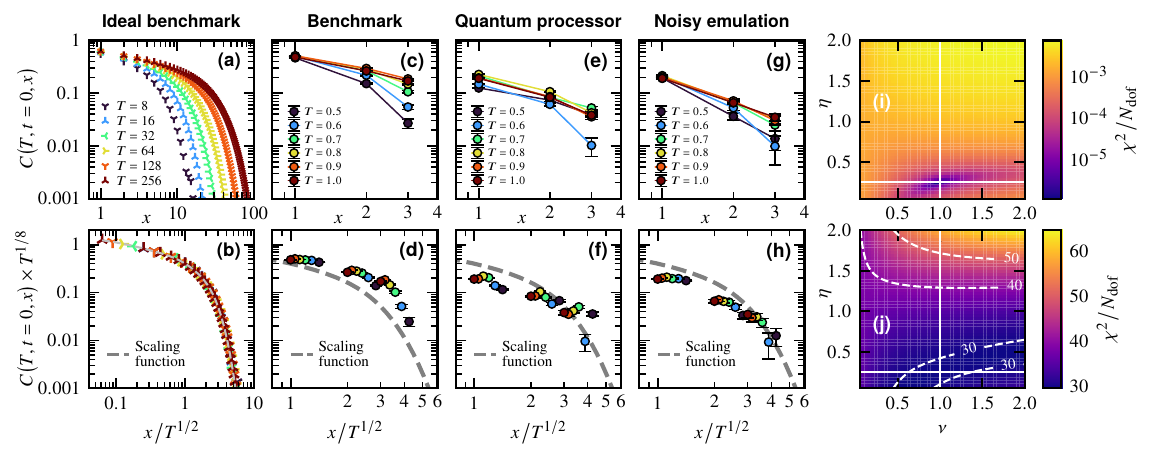}
    \caption{(a)--(c)--(e)--(g) Two-point correlation function of Eq.~\eqref{eq:correlation} at $t=0$ plotted versus the distance $x$ for different drive times $T$. (b)--(d)--(f)--(h) Rescaled two-point correlation function according to Eq.~\eqref{eq:correlation_critical_scaling} with $\nu=z=1$ and $\eta=1/4$. (a), (b) Tensor network emulation of the quantum circuit for $L=257$ qubits with a second-order Suzuki-Trotter expansion and time step $\delta t=0.1$. (c), (d) Perfect emulation of the quantum circuit using $L=7$ qubits and performing two time steps of different duration $\delta t$ to access various drive times $T$. (e), (f) Simulation on Rigetti Aspen-9 superconducting quantum chip using the same parameters as (a) and (b). (g), (h) Noisy emulation of the quantum circuit to model the imperfect hardware. (i), (j) Chi-square per degree of freedom $\chi^2/N_\mathrm{dof}$ quantifying the quality of the data collapse for the two-point correlation function of Eq.~\eqref{eq:correlation} as a function of the critical exponents $\nu$ and $\eta$, see Supplemental Material~\cite{supplemental} (smaller is better). The best collapse should be obtained from the genuine values of $\nu$ and $\eta$. The exact values are marked at the intersection of the two bold straight white lines. (i) Using the benchmark data of (a). (j) Using the quantum processor data of (e).}
    \label{fig:collapse}
\end{figure*}

To verify the scaling law of Eq.~\eqref{eq:correlation_critical_scaling} we emulate the quantum circuit corresponding to an open chain of $L=257$ qubits together with a second-order Suzuki-Trotter expansion and time step $\delta t=0.1$ for different values $T=8,16,\ldots 256$. We set $\hbar=1$. Although it is way out of reach for NISQ hardware, it allows us to obtain benchmark data. The emulation is performed using matrix product states, a well-established and efficient tensor network technique for classically simulating one-dimensional quantum systems~\cite{schollwock2011}. The correlation $C(T,t=0,x)$ is plotted in Fig.~\ref{fig:collapse}(a). We proceed to the rescaling of the data using the exactly known value of the critical exponents of the Ising universality class in $1+1$ dimensions: $\nu=z=1$ and $\eta=1/4$~\cite{francesco2012}. The result is displayed in Fig.~\ref{fig:collapse}(b) where an excellent data collapse is found. An important point that we make in the Supplemental Material~\cite{supplemental} is that, by reducing the standards of an ideal simulation: smaller number of qubits, larger time step, lower-order Suzuki-Trotter expansion, and shorter drive times $T$, one is still able to produce reasonable physics that should be accessible on current NISQ devices, see also Figs.~\ref{fig:collapse}(c)--\ref{fig:collapse}(d).

We now run the quantum circuit on a quantum computer. We use Rigetti Aspen-9 superconducting quantum chip and the provided compiler to translate the quantum circuit into the native gate set~\cite{supplemental}. We work with seven qubits, each directly representing one Ising spin of the Hamiltonians~\eqref{eq:ising_terms}. We perform two time steps using a first-order Suzuki-Trotter expansion, and vary its duration $\delta t$ to access different drive times $T$. We collect $32\,768$ basis states as outputs, from which we compute the two-point correlation function of Eq.~\eqref{eq:correlation}. The raw data are shown in Fig.~\ref{fig:collapse}(d). We observe a distinct decay of the correlation with the distance, but there is no clear hierarchy for the different $T$ values, although the smaller ones tend to be generally lower. Note that, unlike the benchmark emulation, the range of available drive times and distances is more restricted. In the corresponding lower panel, we rescale the data according to Eq.~\eqref{eq:correlation_critical_scaling} and plot for comparison the scaling function extracted from the benchmark data of Fig.~\ref{fig:collapse}(b). There is a good qualitative agreement, despite the hardware being imperfect.

By leaving the exponents $\nu$ and $\eta$ as free parameters and solving the optimization problem seeking to maximize the quality of the data collapse (e.g., by minimizing the chi-square per degree of freedom $\chi^2/N_\mathrm{dof}$)~\cite{Sandvik2010,supplemental}, we can extract a region of maximum likelihood for their values. The corresponding results for the benchmark and quantum processor data are shown in Figs.~\ref{fig:collapse}(i) and~\ref{fig:collapse}(j). The procedure on the benchmark data gives back the known values of the critical exponents. As for the experimental data, we are not able to precisely determine values for the exponents, as there is no clear minimum for the chi-square (cause by a smaller number of qubits, a smaller range of drive times $T$, noise, etc.). Nonetheless, we find that the exact values are within the region with minimum $\chi^2/N_\mathrm{dof}$, and which provides bounds for the exponents. We expect that the continuous improvement of NISQ processors will tighten the bound on the exponent values, see the Supplemental Material for additional data~\cite{supplemental}.

Noise is inherent in NISQ devices, has various origins, and is by definition machine-specific. Familiar sources include decoherence through relaxation and dephasing, readout error, and the qubits being imperfect two-level systems, which can result in faulty quantum operations. Here, we model the effect of noise with a depolarizing channel~\cite{Nielsen2011}. The noisy system is emulated by performing the following stochastic modification to the quantum logic gates~\cite{PhysRevLett.52.1657,PhysRevLett.68.580},
\begin{equation}
    \Biggl\{
    \begin{array}{rl}
        \hat{G}_m&\rightarrow~\hat{G}_m\hat{\mathfrak{a}}_m,\\
        \hat{G}_{m,n}&\rightarrow~\hat{G}_{m,n}\bigl(\hat{\mathfrak{a}}_m\otimes\hat{\mathfrak{b}}_n\bigr),
    \end{array}
    ~~\hat{\mathfrak{a}},\hat{\mathfrak{b}}\in\left\{\hat{I},\hat{X},\hat{Y},\hat{Z}\right\},
    \label{eq:noise_model}
\end{equation}
where $\hat{G}_m$ and $\hat{G}_{m,n}$ represent a one-qubit acting on $m$ and a two-qubit gate acting on $(m,n)$, respectively. The probability that it remains unchanged, i.e., $\hat{\mathfrak{a}}(=\hat{\mathfrak{b}})=\hat{I}$, is $1-p$, with $p$ a parameter controlling the strength of noise. All other combinations are uniformly distributed with probabilities $p/3$ and $p/15$ for one- and two-qubit gates, respectively. The process has to be repeated many times to generate random disordered circuits over which the results are averaged. The $\hat{Z}$ gates induce dephasing, the $\hat{X}$ gates induce a qubit flip, and $\hat{Y}$ a mix of the two.

To assess the reasonableness of the noise model of Eq.~\eqref{eq:noise_model}, we emulate the experiment in the presence of noise and attempt to find a value for the parameter $p$, reproducing at best the experimental data of Figs.~\ref{fig:collapse}(e) and~\ref{fig:collapse}(f). Because all the circuits run on the quantum processor involve two time steps using a first-order Suzuki-Trotter expansion, they all have the same form and size: we anticipate that different quantum circuits performing the same task (following, e.g., compilation) will simply lead to a rescaled value of the phenomenological parameter $p$. To that end, we emulate with noise the circuit of Fig.~\ref{fig:introduction}(c). We report the results in Figs.~\ref{fig:collapse}(g) and~\ref{fig:collapse}(h) for $p\approx 0.08$. Despite being a simple one-parameter phenomenological model which may not capture the various imperfections of the hardware, a good agreement with the experimental data is observed, thus validating to some extent the model.

\begin{figure}[t]
    \includegraphics[width=0.95\columnwidth]{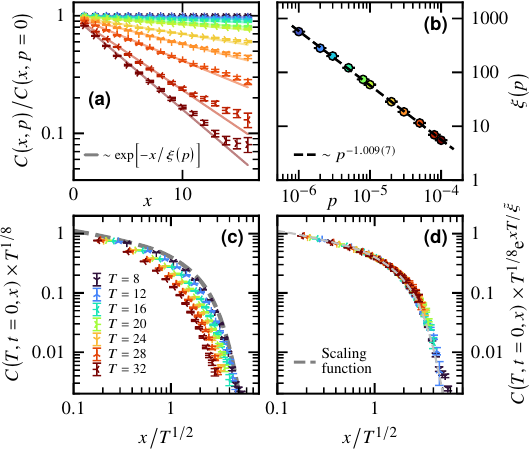}
    \caption{(a) Two-point correlation of Eq.~\eqref{eq:correlation} as a function of the distance $x$, rescaled by the $p=0$ data for $L=33$ qubits. Emulation details: $36\,513$ gates comprised of Hadamard, $\hat{U}^x$, and $\hat{U}^{zz}$ with $T=32$, $t=0$, and second-order Suzuki-Trotter expansion where $\delta t=0.1$. The results are averaged over $\gtrsim 2\times 10^3$ random circuits. Each curve corresponds to a value of $p$ whose code color can be read from panel (b). Fit of the observed exponential decay $\sim\exp\bigl[-x/\xi(p)\bigr]$ (bold line) to extract length $\xi(p)$. (b) Length $\xi(p)$ as a function of $p$, which shows a $\sim 1/p$ dependence. (c), (d) Rescaled two-point correlation function according to Eq.~\eqref{eq:correlation_critical_scaling} ($z=\nu=1$ and $\eta=1/4$). The form of the circuit is the same as the one in panel (a) with $p=10^{-4}$ for various drive times $T$. (d) Same as (c) except that the y-axis is multiplied by an additional term with $\tilde{\xi}\approx 180$, see Eq.~\eqref{eq:correlation_critical_scaling_noise}.}
    \label{fig:noise_nodel}
\end{figure}

To better understand the physics induced by the noise model on the time evolution, we study the combined systems as a function of $p$ using matrix product states. The form and size of the circuit are fixed with $L=33$ qubits and $36\,513$ gates. We plot in Fig.~\ref{fig:noise_nodel}(a) the two-point correlation function of Eq.~\eqref{eq:correlation} rescaled by the noiseless data as a function of the distance $x$. We observe an exponential decay of the form $C\bigl(x,p\bigr)=C\bigl(x,p=0\bigr)\mathrm{e}^{-x/\xi(p)}$, meaning that the noise gives rise to a new length scale in the system. We extract it in Fig.~\ref{fig:noise_nodel}(b) and find that $\xi(p)\sim 1/p$. A simple argument where one supposes that the effect of a single defect in a circuit volume $dx$ (with $d$ the depth) will reduce the correlation by a factor $\varepsilon>1$, leads to $C(x,p)\sim C(x,p=0)/\varepsilon^{pdx}$ for an average number of defects $\sim pdx$, assuming their effect is uncorrelated. It is compatible with the exponential decay observed in the emulations reported in Figs.~\ref{fig:noise_nodel}(a) and~\ref{fig:noise_nodel}(b). The depth dependence $\xi\sim 1/d$ at fixed $p$ is verified in the Supplemental Material~\cite{supplemental}. Note that for a fixed time step $\delta t$, the circuit depth is proportional to the drive time $T$, and we use $d\to T$ in the following. The noise-induced length scale $\xi\sim 1/pT$ competes with the characteristic length scale $\ell\sim T^{\nu/(1+z\nu)}$ of the KZ mechanism. In the one-dimensional quantum Ising model studied here, for the KZ mechanism to dominate over the noise and observe genuine quantum criticality, one needs $p\ll T^{-3/2}\sim L^{-3}$. An analogy can be drawn between the noise in the quantum circuit and thermal effects induced by a finite temperature $\Theta$ in the quantum Ising model, as they both lead to a length scale $\xi^{-1}\sim\Theta\sim p$~\cite{Sachdev1997}. Such an analogy between noise and effective temperature was also reported in open quantum systems~\cite{PhysRevB.93.014307,Paz2019,PhysRevA.104.023713} and sudden quench protocols subject to a time-dependent white noise~\cite{PhysRevB.86.060408,PhysRevB.89.024303}. Interestingly, one can include a new parameter $\tilde{\xi}=T\xi$ in the critical scaling of Eq.~\eqref{eq:correlation_critical_scaling}, accounting for the effect of noise on quantum criticality,
\begin{equation}
    \mathcal{F}\bigl(x/\ell, t/\tau\bigr)~\to~\mathcal{F}\bigl(x/\ell,t/\tau\bigr)\times\exp\bigl(-xT/\tilde{\xi}\bigr).
    \label{eq:correlation_critical_scaling_noise}
\end{equation}
Equation~\eqref{eq:correlation_critical_scaling_noise} is confirmed by emulations based on matrix product state for $L=33$ qubits and $p=10^{-4}$, with the form of the circuit and other parameters similar to those of Figs.~\ref{fig:noise_nodel}(a) and~\ref{fig:noise_nodel}(b). The raw and noise-corrected data collapses are displayed side by side in Figs.~\ref{fig:noise_nodel}(c) and~\ref{fig:noise_nodel}(d), with a substantial improvement upon including the parameter $\tilde{\xi}\approx 180$, which can be found without any prior knowledge, comparably to the critical exponents~\cite{supplemental}. The reduced connectivity at the boundaries of the system makes the exponential decay of Fig.~\ref{fig:noise_nodel}(a) drifts for these qubits, and the noise correction is not directly applicable on smaller-scale systems, such as the ones simulated on the quantum processor displayed in Figs.~\ref{fig:collapse}(e) and~\ref{fig:collapse}(f).

While quantum criticality is well-understood in $1+1$ dimensions, much less is known beyond that. The absence of efficient classical methods to simulate certain types of quantum many-body systems, e.g., interacting fermions or frustrated magnets, limits our microscopic understanding of these phases of matter and their transitions. Here, we have shown that current NISQ devices can simulate quantum criticality by leveraging a dynamically-driven phenomenon. Using a programmable superconducting processor, we demonstrated this approach on the one-dimensional quantum Ising model by obtaining a good agreement with benchmark data. Despite the limited number of qubits and the restricted depth of the quantum circuits, we estimated the critical exponents. The continuous improvement of NISQ hardware will generate better quality and larger-scale data. Not only will it leads to more accurate results, but it will also open the way to uncharted problems. In addition, we have shown that one can directly account for the inherent noise of the current generation of quantum computers. We found that the noise induces a length scale controlling how far qubits can be nontrivially correlated. It can be included in scaling laws, thus making the noise irrelevant to some extent when investigating quantum criticality. Whether this noise-induced length scale is a general feature arising in other quantum algorithms remains to be explored, as similar behavior was recently reported in other kinds of many-body problems~\cite{PRXQuantum.2.030346,PhysRevLett.126.230501}.

\let\oldaddcontentsline\addcontentsline
\renewcommand{\addcontentsline}[3]{}
\begin{acknowledgments}
    We gratefully acknowledge the support of B. Evert, M. Hodson, M. Paini, and M.J. Reagor at Rigetti Computing. M.D. also acknowledges discussions with A. Avdoshkin, E.G. Dalla Torre, F. Machado, T. Scaffidi, and N.E. Sherman. We thank one of the anonymous referees for suggesting to us to consider the strong noise regime~\cite{supplemental}. M.D. received support from the U.S. Department of Energy, Office of Science, Office of Basic Energy Sciences, Materials Sciences and Engineering Division under Award No. DE-AC02-05-CH11231 through the Theory Institute for Materials and Energy Spectroscopy (TIMES). J.E.M. was supported by the Quantum Science Center (QSC), a National Quantum Information Science Research Center of the U.S. Department of Energy (DOE), and a Simons Investigatorship. This research used the Lawrencium computational cluster resource provided by the IT Division at the Lawrence Berkeley National Laboratory (supported by the Director, Office of Science, Office of Basic Energy Sciences, of the U.S. Department of Energy under Award No. DE-AC02-05CH11231). This research also used resources of the National Energy Research Scientific Computing Center (NERSC), a U.S. Department of Energy Office of Science User Facility operated under Award No. DE-AC02-05CH11231. In addition, this research used resources of the Oak Ridge Leadership Computing Facility at the Oak Ridge National Laboratory, which is supported by the Office of Science of the U.S. Department of Energy under Award No. DE-AC05-00OR22725.
\end{acknowledgments}
\let\addcontentsline\oldaddcontentsline

\let\oldaddcontentsline\addcontentsline
\renewcommand{\addcontentsline}[3]{}
\bibliography{references}
\let\addcontentsline\oldaddcontentsline

\newpage\cleardoublepage

\onecolumngrid
\setcounter{page}{1}
\setcounter{secnumdepth}{3}
\setcounter{figure}{0}
\setcounter{equation}{0}
\renewcommand\thefigure{S\arabic{figure}}
\renewcommand\theequation{S\arabic{equation}}

\setlength{\belowcaptionskip}{0pt}

\begin{center}
    \large\textbf{Supplemental Material for\\``\textit{Quantum criticality using a superconducting quantum processor}''}
\end{center}

\addvspace{5mm}
\begin{center}
    \begin{minipage}{0.8\textwidth}
        We provide more information on the quantum circuits implementation, the emulation techniques, additional emulation data studying the effect of the different parameters, details on how the critical exponents are extracted, additional data on the noise model, and details regarding the quantum hardware.
    \end{minipage}
\end{center}

{\hypersetup{linkcolor=black}\tableofcontents}
\newpage

\section{Quantum circuit}

\subsection{Quantum logic gates}

For completeness, we recall the matrix form of the different quantum logic gates used and/or mentioned throughout this work. They are given in the standard computational $z$-basis. The one-qubit gates are,
\begin{equation}
    \begin{quantikz}
        & \gate{X} & \qw
    \end{quantikz}
    = \begin{pmatrix} 0 & 1 \\ 1 & 0 \end{pmatrix},
    \begin{quantikz}
        & \gate{Y} & \qw
    \end{quantikz}
    = \begin{pmatrix} 0 & -i \\ i & 0 \end{pmatrix},
    \begin{quantikz}
        & \gate{Z} & \qw
    \end{quantikz}
    = \begin{pmatrix} 1 & 0 \\ 0 & -1 \end{pmatrix},
    \begin{quantikz}
        & \gate{I} & \qw
    \end{quantikz}
    = \begin{pmatrix} 1 & 0 \\ 0 & 1 \end{pmatrix},
\end{equation}
\begin{equation}
    \begin{quantikz}
        & \gate{H} & \qw
    \end{quantikz}
    = \frac{1}{\sqrt{2}} \begin{pmatrix} 1 & 1 \\ 1 & -1 \end{pmatrix},
    \begin{quantikz}
        & \gate{R_z(\phi)} & \qw
    \end{quantikz}
    = \begin{pmatrix} \mathrm{e}^{-i\phi/2} & 0 \\0 & \mathrm{e}^{+i\phi/2}\end{pmatrix},
    \begin{quantikz}
        & \gate{R_x(\phi)} & \qw
    \end{quantikz}
    = \begin{pmatrix} \cos(\phi/2) & -i\sin(\phi/2) \\ -i\sin(\phi/2) & \cos(\phi/2) \end{pmatrix},
\end{equation}
with the Pauli operators, the Identity matrix, the Hadamard gate, the rotation gate around the $z$ axis, and the rotation gate around the $x$ axis. The two-qubit gates are,
\begin{equation}
    \begin{quantikz}\textsf{CNOT}~=\end{quantikz}
    \begin{quantikz}
        & \ctrl{1} & \qw\\
        & \targ & \qw & \qw
    \end{quantikz}
    = \begin{pmatrix} 1 & 0 & 0 & 0 \\ 0 & 1 & 0 & 0 \\ 0 & 0 & 0 & 1 \\ 0 & 0 & 1 & 0 \end{pmatrix},
    \begin{quantikz}\textsf{CPHASE}(\phi)~=\end{quantikz}
    \begin{pmatrix} 1 & 0 & 0 & 0 \\ 0 & 1 & 0 & 0 \\ 0 & 0 & 1 & 0 \\ 0 & 0 & 0 & \mathrm{e^{i\phi}} \end{pmatrix}
    \begin{quantikz},~~~\textsf{CZ}\equiv\textsf{CPHASE}(\phi=\pi)\end{quantikz}
    = \begin{quantikz}
        & \ctrl{1} & \qw\\
        & \control{} & \qw
    \end{quantikz},
\end{equation}

\subsection{Implementation of the time-evolution}

We employ the same notation as in the main text, and provide additional information on the construction of the quantum circuits. We first get the initial state $\vert\Psi(t=-T)\rangle$, corresponding to the ground state of the the paramagnetic term $-\sum_{n=1}^L\hat{X}_n$ of the quantum Ising model, by applying Hadamard gates on the individual $L$ qubits. Then, the time evolution of the quantum state requires translating the piecewise constant time-evolution operator $\exp\bigl[-i\hat{\mathcal{H}}(T,t)\delta t\bigr]$ into a quantum circuit. We employ a Suzuki-Trotter expansion~\cite{hatano2005} which allows us to break down the evolution operator acting onto $L$ qubits into smaller pieces acting on at most two qubits. Because not all terms of the Hamiltonian $\hat{\mathcal{H}}(T,t)$ commute with one another due to the intrinsic commutation relations of Pauli operators, the Suzuki-Trotter expansion is not exact and induces a systematic error. At first order, the Suzuki-Trotter expansion reads,
\begin{equation}
    \exp\Bigl[-i\hat{\mathcal{H}}\bigl(T,t\bigr)\delta t\Bigr]\simeq\Biggl[\prod\nolimits_{n=2j,\;j\in\mathbb{N}}^{n\leq L}\hat{U}^{zz}_n\bigl(T,t,\delta t\bigr)\Biggr]\Biggl[\prod\nolimits_{n=2j-1,\;j\in\mathbb{N}}^{n\leq L}\hat{U}^{zz}_m\bigl(T,t,\delta t\bigr)\Biggr]\Biggl[\prod\nolimits_{n=1}^L\hat{U}^x_m\bigl(T,t,\delta t\bigr)\Biggr]+O\left(\delta t^2\right),
    \label{eq:trotter_order1}
\end{equation}
with,
\begin{equation}
    \hat{U}^x_n\bigl(T,t,\delta t\bigr)=\exp\left[i\delta t\left(1-\frac{t}{T}\right)\hat{X}_n\right],~~\mathrm{and}~~\hat{U}^{zz}_n\bigl(T,t,\delta t\bigr) = \exp\left[i\delta t\left(1+\frac{t}{T}\right)\hat{Z}_n\hat{Z}_{n+1}\right].
\end{equation}
Both are easily translated into standard quantum logic gates. The former is directly related to a single-qubit rotation gate around the $x$ axis: $R_x(\phi)$ with $\phi=-2\delta t(1-t/T)$. The latter can be decomposed by sandwiching a single-qubit rotation gate around the $z$ axis $R_z(\phi)$ by two-qubit \textsf{CNOT} gates where $\phi=-2\delta t(1+t/T)$, see Fig.~1(b) in the main text.

In practice, the two-qubit gates $\hat{U}^{zz}_n(T,t,\delta t)$ can first be simultaneously applied on even bonds and then simultaneously on odd bonds, thus limiting the circuit depth. Higher order expansions yielding a smaller error can be constructed at the price of additional time-evolution operators in the expansion. For instance, by symmetrizing Eq.~\eqref{eq:trotter_order1} some of the errors cancel and one obtains a second-order Suzuki-Trotter expansion in the form,
\begin{equation}
    \begin{split}
        \exp\Bigl[-i\hat{\mathcal{H}}\bigl(T,t\bigr)\delta t\Bigr]\simeq&\quad\Biggl[\prod\nolimits_{n=1}^L\hat{U}^x_n\left(T,t,\frac{\delta t}{2}\right)\Biggr]\Biggl[\prod\nolimits_{n=2j,\;j\in\mathbb{N}}^{n\leq L}\hat{U}^{zz}_n\left(T,t,\frac{\delta t}{2}\right)\Biggr]\Biggl[\prod\nolimits_{n=2j-1,\;j\in\mathbb{N}}^{n\leq L}\hat{U}^{zz}_n\bigl(T,t,\delta t\bigr)\Biggr]\\
        &\times\Biggl[\prod\nolimits_{n=2j,\;j\in\mathbb{N}}^{n\leq L}\hat{U}^{zz}_n\left(T,t,\frac{\delta t}{2}\right)\Biggr]\Biggl[\prod\nolimits_{n=1}^L\hat{U}^x_n\left(T,t,\frac{\delta t}{2}\right)\Biggr] + O\left(\delta t^3\right).
    \end{split}
    \label{eq:trotter_order2}
\end{equation}

\subsection{Measurements}

We perform measurements directly in the computational $z$-basis by measuring the state of each qubit at the end of the quantum circuit. For each circuit instance being run, we get a basis state as the output, e.g., $\vert\boldsymbol{\sigma}^{[n]}\rangle\equiv\vert011\cdots 0\rangle$. For a total of $n=1,2\ldots N$ outputs, the two-point correlation between the qubits $i$ and $j$, which is a diagonal observable in the computational basis, reads,
\begin{equation}
    \Bigl\langle\hat{Z}_i\hat{Z}_j\Bigr\rangle\approx \frac{4}{N}\sum\nolimits_{n=1}^N\left(\sigma_i^{[n]}-\frac{1}{2}\right)\left(\sigma_j^{[n]}-\frac{1}{2}\right),
    \label{eq:mes_correlation}
\end{equation}
with $\sigma_i^{[n]}$ the value ($0$ or $1$) of the measured qubit $i$ at the end of the run $n$. The finite number of outputs to compute Eq.~\eqref{eq:mes_correlation} results in error bars vanishing as $\sim 1/\sqrt{N}$. In practice, we select $i$ as the reference qubit located in the middle of the system, and have $j=r\pm x$, with $x$ the distance between $i$ and $j$, i.e., $x=\vert i-j\vert$. For a system symmetric around the reference qubit $r$, we expect the correlations measured at $\pm x$ to be the identical. Thus, the two-point correlation $C(T,t,x)$ considered here and in the main text, corresponds to the average over $\pm x$.

\subsection{Practical implementation on quantum hardware}

The native set of quantum logic gates available on the Rigetti Aspen-9 chip includes the single-qubit rotation $R_z(\phi)$ around the $z$ axis, the single-qubit rotation around the $x$ axis $R_x(\mathbb{Z}\pi/2)$, the two-qubit \textsf{CPHASE}, \textsf{CZ}, and \textsf{XY} gates. To obtain the Hadamard gate, one can use the following decomposition,
\begin{equation}
    \begin{quantikz}
        & \gate{H} & \qw
    \end{quantikz}
    \begin{quantikz}=\end{quantikz}
    \begin{quantikz}
        & \gate{R_z(\pi/2)} & \gate{R_x(\pi/2)} & \gate{R_z(\pi/2)} & \qw
    \end{quantikz}
    \label{eq:hadamard_decomposition}
\end{equation}
The rotation around the $x$ axis for an arbitrary angle $\phi$ (not limited to multiples of $\pi/2$) can be implemented as,
\begin{equation}
    \begin{quantikz}
        & \gate{R_x(\phi)} & \qw
    \end{quantikz}
    \begin{quantikz}=\end{quantikz}
    \begin{quantikz}
        & \gate{H} & \gate{R_z(\phi)} & \gate{H} & \qw
    \end{quantikz}
\end{equation}
with the Hadamard gate following the decomposition of Eq.~\eqref{eq:hadamard_decomposition}. Finally, the \textsf{CNOT} gate is obtained by means of a \textsf{CZ} gate,
\begin{equation}
    \begin{quantikz}
        & \ctrl{1} & \qw\\
        & \targ & \qw & \qw
    \end{quantikz}
    \begin{quantikz}=\end{quantikz}
    \begin{quantikz}
        & \qw & \ctrl{1} & \qw & \qw\\
        & \gate{H} & \control{} & \gate{H} & \qw
    \end{quantikz}
\end{equation}
These operations are typically taken care of by a compiler. Some of these additional gates may be simplified in the final quantum circuit, e.g.,
\begin{equation}
    \begin{quantikz}
        & \gate{R_z(\varphi)} & \gate{R_z(\phi)} & \qw
    \end{quantikz}
    \begin{quantikz}=\end{quantikz}
    \begin{quantikz}
        & \gate{R_z(\varphi+\phi)} & \qw
    \end{quantikz},
     \begin{quantikz}~\mathrm{and}~\end{quantikz}
    \begin{quantikz}
        & \gate{H} & \gate{H} & \qw
    \end{quantikz}
    \begin{quantikz}=\end{quantikz}
    \begin{quantikz}
        & \gate{I} & \qw
    \end{quantikz}
    \begin{quantikz}(\mathrm{Identity})\end{quantikz}
\end{equation}

\subsection{Emulation}

\subsubsection{Matrix product states}

A matrix product state is a well-established and efficient type of tensor network for classically simulating one-dimensional quantum systems~\cite{schollwock2011}. Our calculations are based on the time-evolving block decimation algorithm~\cite{PhysRevLett.93.040502}, implemented using the ITensor library~\cite{itensor} together with a maximum bond dimension of $1\,024$ and a cutoff of $10^{-12}$ when performing the singular value decompositions. When using matrix product states, the measurements are exactly performed by summing over all the basis states, and there is no statistical error associated to a finite number of outputs.

\subsubsection{Conventional quantum circuit emulation}

In addition to matrix product states, we have implemented smaller-scale quantum circuits independently on two different platforms: Cirq~\cite{cirq} supplemented by qsim~\cite{qsim}, and PyQuil~\cite{pyquil}. We used Cirq to generate the main results presented in this work---other than those obtained with matrix product states, which are explicitly stated---and we used PyQuil to interface with Rigetti quantum processor. Unlike the matrix product state data, the results are averaged over a finite number of outputs, as in a realistic quantum computer run, leading to a statistical error.

\section{Critical scaling away from ideality}

The goal of this section is to show that the critical scaling reported as a benchmark in the main text, for $L=257$ qubits together with a second-order Suzuki-Trotter expansion and time step $\delta t=0.1$ for different drive time values $T=8,16,\ldots 256$, can also be observed on smaller-scale circuits.  Understanding the dependence of critical scaling on circuit size is necessary as $L=257$ is well out of reach for current NISQ processors.

\subsection{System size and drive time}

Here, we study the effect of the system size $L$ and the drive time $T$. A perfect setup would require $L\to+\infty$ degrees of freedom and then $T\to+\infty$ to be in the adiabatic limit. In this section, we show that genuine physics can be obtained away from these ideal limits by looking at the two-point correlation $C(T,t,x)$, defined in the main text.

\subsubsection{Spatiotemporal dependence of the two-point correlation}

\begin{figure}[h]
	\centering
	\includegraphics[width=0.6\textwidth]{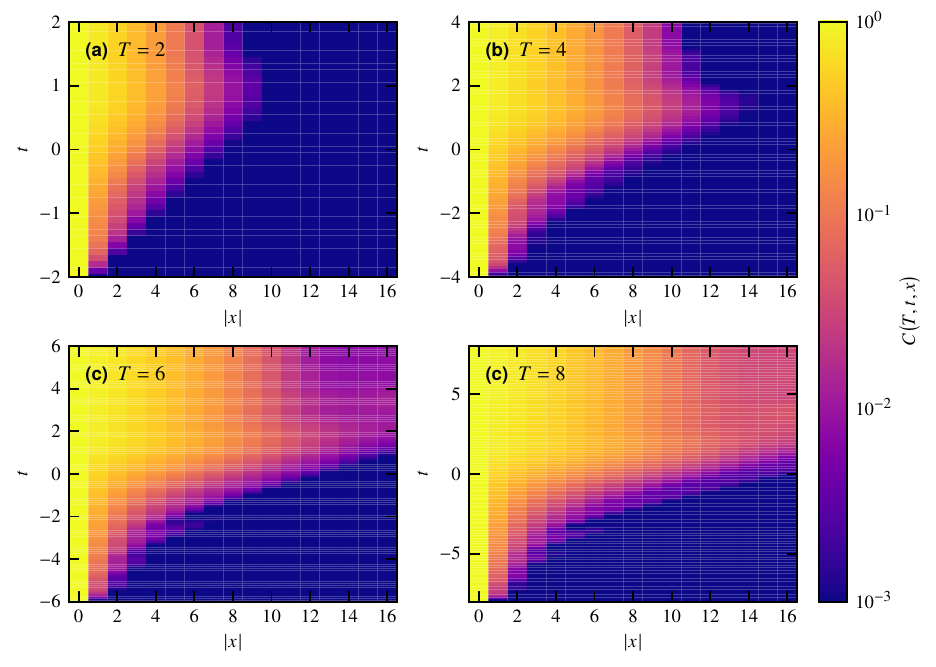}
	\caption{Log-scale intensity plot of the two-point correlation $C(T,t,x)$ measured for $L=33$ and various drive times $T$: (a) $T=2$, (b) $T=4$, (c) $T=6$, and (d) $T=8$. The data were generated using matrix product states with a second-order Suzuki-Trotter expansion and time step $\delta t=0.1$. There is a cutoff for the intensity values below $10^{-3}$.}
	\label{fig:intensity_plots}
\end{figure}

First, we consider a fixed size $L=33$ for different values of the drive time $T$ and show the corresponding intensity plot of the two-point correlation $C(T,t,x)$ in Fig.~\ref{fig:intensity_plots}. As $T$ increases, the correlation between degrees of freedom builds up to longer and longer distances $x$. We also observe that once the QCP is crossed at $t=0$, the correlation stops expanding spatially. As long as $C(T,t,x)\to 0$ close to the boundary of the system, one can consider the results as being free of finite-size effects. In other words, one would get similar data for larger values of $L$. The characteristic length scale $\ell$ of the Kibble--Zurek mechanism goes as $\ell\sim T^{\nu/(1+z\nu)}\sim\sqrt{T}$ (using $z=\nu=1$ for the Ising universality class in $1+1$ dimensions). Substituting $\ell\to L$, one finds that the minimum required system size goes as $L\sim\sqrt{T}$. Here, we find from Fig.~\ref{fig:intensity_plots} that $L=33$ is large enough for $T\lesssim 4$ if one wants to access times up to $t=+T$.

\subsubsection{Critical scaling on small system sizes}

\begin{figure}[h]
	\centering
	\includegraphics[width=0.6\textwidth]{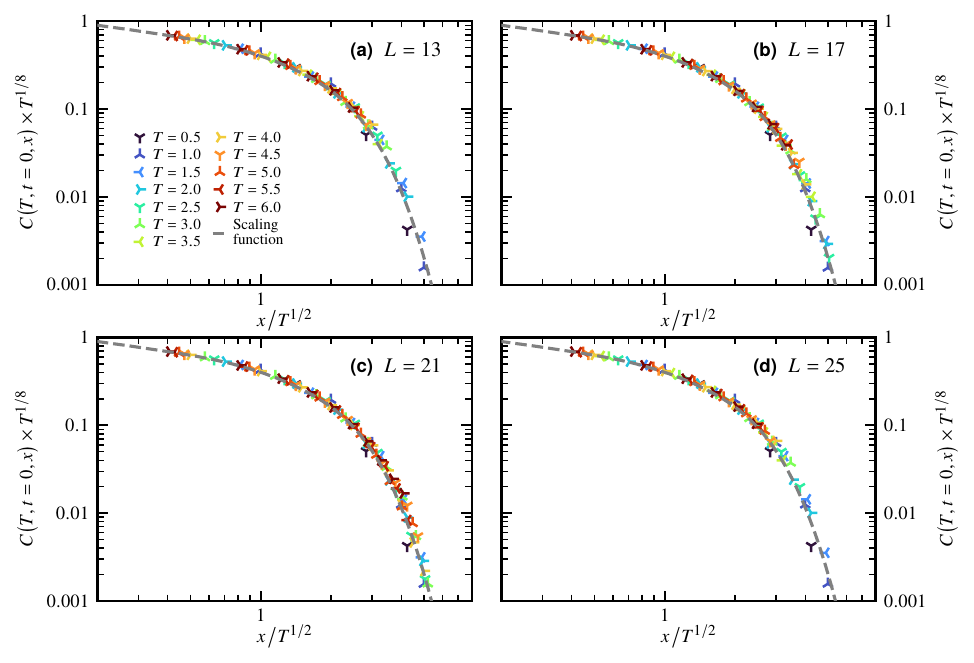}
	\caption{Data collapse following the quantum critical scaling of the two-point correlation $C(T,t=0,x)$ measured for different system sizes $L$ and various drive times $T$ at $t=0$ corresponding to the QCP. The data were generated using matrix product states with a second-order Suzuki-Trotter expansion and time step $\delta t=0.1$. (a) $L=13$, (b) $L=17$, (c) $L=21$, and (d) $L=25$. The scaling function (dashed line) is from the benchmark data of the main text.}
	\label{fig:collapse_small_sizes}
\end{figure}

We consider the quantum critical scaling of the two-point correlation $C(T,t=0,x)$ for small system sizes, from $L=13$ to $L=25$. We rescale $C(T,t=0,x)\to C(T,t=0,x)\times T^{\nu\eta/(1+z\nu)}$ and $x\to x/T^{\nu/(1+z\nu)}$, and plot the data in Fig.~\ref{fig:collapse_small_sizes}. We consider the Ising universality class in $1+1$ dimensions with critical exponents $\nu=z=1$ and $\eta=1/4$, giving $\nu\eta/(1+z\nu)=1/8$ and $\nu/(1+z\nu)=1/2$. We find a very good collapse, and a very good agreement with the benchmark scaling function obtained on much larger system sizes and much longer drive times, see Figs.~2(a)--(b) in the main text.

\subsubsection{Critical scaling on small system sizes with short drive times}

\begin{figure}[h]
	\centering
	\includegraphics[width=0.6\textwidth]{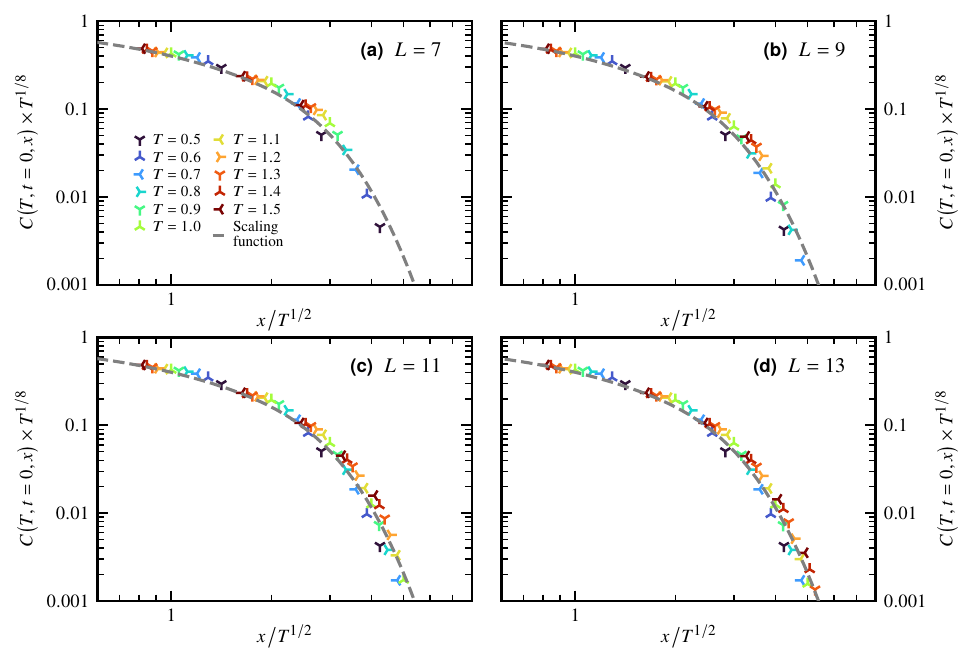}
	\caption{Data collapse following the quantum critical scaling of the two-point correlation $C(T,t=0,x)$ measured for different system sizes $L$ and various drive times $T$ at $t=0$ corresponding to the QCP. The data were generated using matrix product states with a second-order Suzuki-Trotter expansion and time step $\delta t=0.1$. (a) $L=7$, (b) $L=9$, (c) $L=11$, and (d) $L=13$. The scaling function (dashed line) is from the benchmark data of the main text.}
	\label{fig:collapse_small_sizes_small_drives}
\end{figure}

The number of qubits involved in the data of Fig.~\ref{fig:collapse_small_sizes} is accessible on NISQ hardware. However, the circuit depth is still too large. One way to reduce it is by considering smaller drive times $T$. We plot the quantum critical scaling of the two-point correlation $C(T,t=0,x)$ in Fig.~\ref{fig:collapse_small_sizes_small_drives}, for system sizes ranging from $L=7$ to $L=13$, and drive times $T\in[0.5,1.5]$. We note that although the data collapse does not perfectly align with the benchmark scaling function (as it is the case for larger system sizes and longer drive times $T$), there is still a reasonable agreement.

\subsection{Suzuki-Trotter expansion error}

\begin{figure}[!t]
	\centering
	\includegraphics[width=0.65\textwidth]{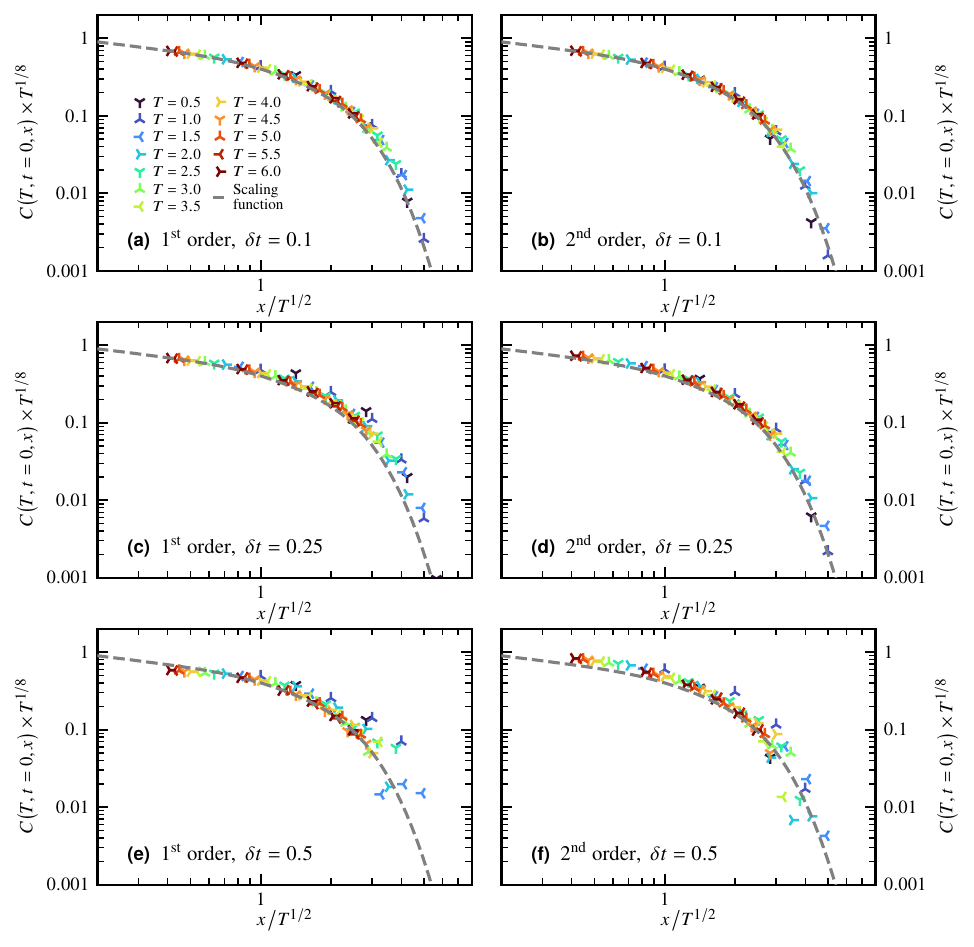}
	\caption{Data collapse following the quantum critical scaling of the two-point correlation $C(T,t=0,x)$ measured for $L=13$ and various drive times $T$ at $t=0$ corresponding to the QCP. The data were generated using matrix product states. The different panels correspond to different orders of the Suzuki-Trotter expansion and time steps $\delta t$: (a) $1^\mathrm{st}$~order, $\delta t=0.1$, (b) $2^\mathrm{nd}$~order, $\delta t=0.1$, (c) $1^\mathrm{st}$~order, $\delta t=0.25$, (d) $2^\mathrm{nd}$~order, $\delta t=0.25$, (e) $1^\mathrm{st}$~order, $\delta t=0.5$, (f) and $2^\mathrm{nd}$~order, $\delta t=0.5$. The scaling function (dashed line) is from the benchmark data of the main text.}
	\label{fig:trotter_error}
\end{figure}

Another way to reduce the circuit depth when performing the time evolution is by reducing the order of the Suzuki-Trotter expansion and/or increasing the time step $\delta t$, at the price of a systematic error. A Suzuki-Trotter expansion of order $\kappa$ introduces a systematic error $O\bigl(\delta t^{\kappa+1}\bigr)$, see Eqs.~\eqref{eq:trotter_order1}--\eqref{eq:trotter_order2}. Moreover, the Suzuki-Trotter expansion with $\kappa=2$ has a larger circuit depth than with $\kappa=1$. As we want to minimize the overall circuit depth for NISQ hardware while accessing long drive values $T$, it is instructive to study the effect of the time step $\delta t$ in combination with the Suzuki-Trotter expansion order $\kappa$. Here, we study the effect of these two parameters on the quality of the data collapse. We compute the two-point correlation $C(T,t=0,x)$ for $L=13$ at $t=0$ for different values of the drive time $T$.

We plot in Fig.~\ref{fig:trotter_error} the data collapse for $\kappa=1$, $\kappa=2$ and for time steps $\delta t=0.1$, $0.25$, and $0.5$. While the quality of the collapse is systematically better for $\kappa=2>\kappa=1$ and $\delta t=0.1<0.25<0.5$, one still obtains good results for $\kappa=1$ and $\delta t=0.5$, corresponding to the smallest circuit depth of all the pairs of parameters. Note that for $\delta t=0.5$, the data for the drive time $T=0.5$ corresponds to a single step in the time evolution. This explains why the $T=0.5$ data are off further from the rest of the collapse.

\subsection{Effect of the finite number of outputs}

\begin{figure}[t]
	\centering
	\includegraphics[width=0.6\textwidth]{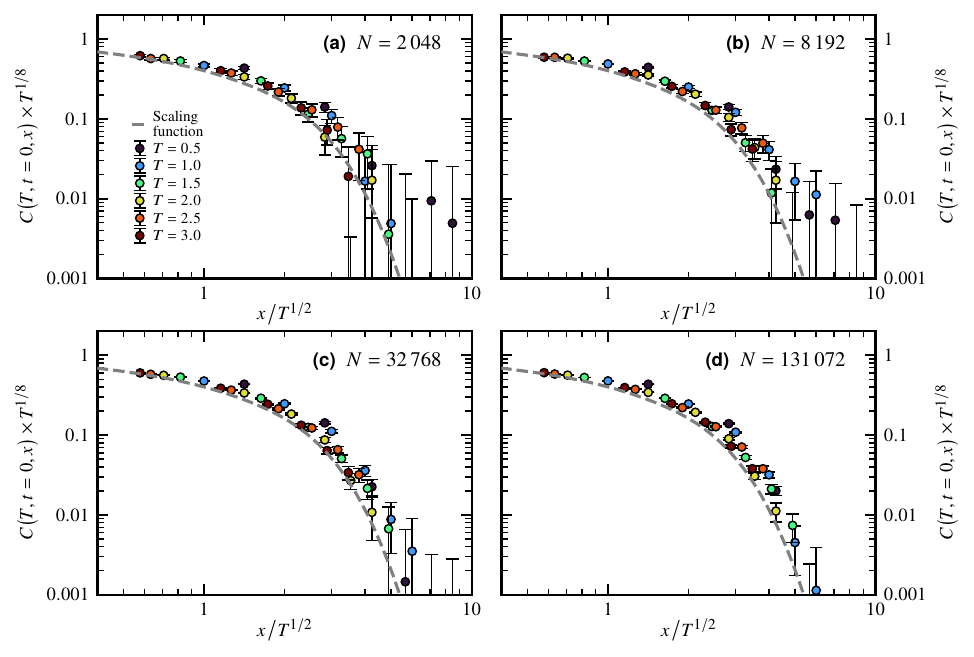}
	\caption{Data collapse following the quantum critical scaling of the two-point correlation $C(T,t=0,x)$ measured for $L=13$ and various drive times $T$ at $t=0$ corresponding to the QCP. The different panels correspond to a different number of outputs $N$ used to compute the two-point correlation: (a) $N=2\,048$, (b) $N=8\,192$, (c) $N=32\,768$, and (d) $N=131\,072$. The scaling function (dashed line) is from the benchmark data of the main text.}
	\label{fig:number_of_outputs}
\end{figure}

The number of outputs $N$ used to compute physical observables corresponds to the number of times the quantum circuit is executed. Here, we study the effect of the number of outputs N on the quality of the data collapse. We set $L=13$ and use a first-order Suzuki-Trotter expansion together with a time step $\delta t=0.25$. We compute the data collapse of the two-point correlation $C(T,t=0,x)$ for various number of outputs $N$ ranging from $2^{11}$ to $2^{17}$, see Fig.~\ref{fig:number_of_outputs}. As expected, the resolution gets much better as $N$ increases and the error bars gets reduced accordingly. This is the result of the correlation $C(T,t=0,x)$ decaying with $x$, leading to smaller values $C(T,t=0,x)\to 0$ as $x$ increases. In order to resolve these values, one needs a larger number of outputs, with the resolution improving as $\sim\sqrt{N}$.

\section{Determining the critical exponents}

\subsection{Method}

We seek to find the numerical value of the critical exponents entering the critical scaling relations assuming they are unknown. One way to achieve that is by casting this problem as an optimization problem, which seeks to maximize the quality of the data collapse, with the best collapse obtained for the genuine values of the critical exponents.

We consider the two-point correlation function with the scaling law of the main text. We set $z=1$ (as it is often the case in quantum phase transitions), leaving two exponents to be determined: the correlation length exponent $\nu$ and the anomalous exponent $\eta$. To do so, we express the (unknown) scaling function $\mathcal{F}$ through a Taylor expansion times an exponential component which accounts for the rapid decay of the two-point correlation, see benchmark data of Figs.~2(a)--(b) in the main text,
\begin{equation}
    Y\bigl(\nu,\eta\bigr) = \mathcal{F}\left[X\bigl(\nu,\eta\bigr)\right]\approx\left[\sum\nolimits_{m=0}^Ma_m X^m\bigl(\nu,\eta\bigr)\right]\times\exp\Bigl[-\tilde{a}X\bigl(\nu,\eta\bigr)\Bigr],
    \label{eq:fit_exponents}
\end{equation}
with $Y\bigl(\nu,\eta\bigr)=C\bigl(T,t,x\bigr)\times T^{\nu\eta/(1+z\nu)}$ and $X\bigl(\nu,\eta\bigr)=x\times T^{-\nu/(1+z\nu)}$ according to the scaling law of the two-point correlation function. The order of the Taylor expansion $M$ is a parameter. For given values of $\nu$ and $\eta$, we perform a least-square fitting of the data with parameters $\tilde{a}$, $a_0$, $a_1$\ldots, $a_M$. Assuming $N$ pairs of data points $\bigl\{X_i,Y_i\bigr\}$, the quality of the fit is measured from the chi-squared statistic,
\begin{equation}
    \frac{\chi^2\bigl(\nu,\eta\bigr)}{N_\mathrm{dof}}=\frac{1}{N_\mathrm{dof}}\sum\nolimits_{i=1}^N\left(\frac{Y_i\bigl(\nu,\eta\bigr)-\mathcal{F}\left[X_i\bigl(\nu,\eta\bigr)\right]}{\Delta Y_i\bigl(\nu,\eta\bigr)}\right)^2,
    \label{eq:chi2}
\end{equation}
with $N_\mathrm{dof}=(N-\mathrm{number~of~free~parameters})$, and with $\Delta Y_i\bigl(\nu,\eta\bigr)$ the error on $Y_i\bigl(\nu,\eta\bigr)$. By repeating this procedure for various values of $\nu$ and $\eta$, we can estimate the best values for the critical exponents when $\chi^2\bigl(\nu,\eta\bigr)/N_\mathrm{dof}$ is minimized.

\subsection{Application to the data of the main text}

This procedure is used in Figs.~2(g)--(h) of the main text. For the benchmark data, because there is no error on $Y_i$, all data points are equally weighted in the fit by setting $\Delta Y_i=1$. We Taylor expand the scaling function of Eq.~\eqref{eq:fit_exponents} to $M=4$. We find that the best collapse is obtained for $\nu\simeq 0.98$ and $\eta\simeq 0.25$. It is in very good agreement with the exact values of the critical exponents. The same procedure is performed on the experimental data, with $M=4$, but by forcing the parameter of the exponential component to zero ($\tilde{a}=0$), as we have found that we cannot obtain reliable fits when it is included---unlike the benchmark data, the quantum processor simulation data are only available in a smaller window because of the smaller system size $L=7$ and drive times $T\leq 1$. We are not able to precisely determine values for the exponents, as there is no clear minimum for the chi-square. Nonetheless, we find that the exact values are within the region with minimum $\chi^2/N_\mathrm{dof}$, and which provides crude bounds for the exponents. We expect that the continuous improvement of NISQ processors will tighten such bounds on the exponent values, see next section.

\subsection{Application to intermediate-scale noisy emulations}

\begin{figure}[h]
	\centering
	\includegraphics[width=0.6\textwidth]{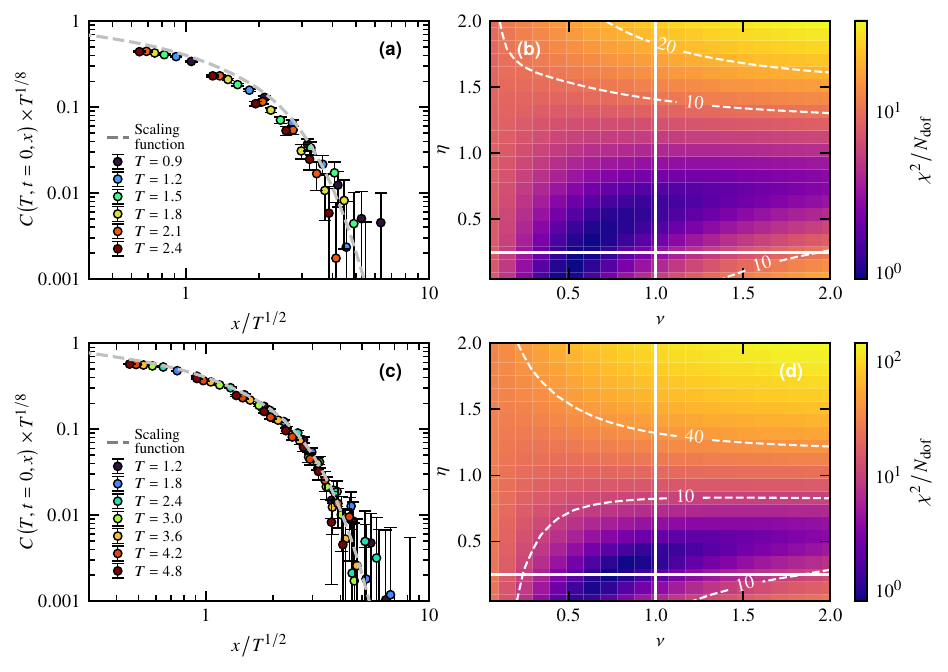}
	\caption{Left column: Data collapse following the quantum critical scaling of the two-point correlation $C(T,t=0,x)$ measured for (a) $L=13$ with $6$ time steps and (b) $L=19$ with $12$ time steps by varying $\delta t$ to access different drive times $T$. A first-order Suzuki-Trotter decomposition is considered. The data is emulated from the circuit of Fig.~1(c) in the main text with a noise parameter (a) $p=0.01$ and (b) $p=0.002$. We collect $32\,768$ basis states as outputs to compute averages. Right column: Chi-square per degree of freedom $\chi^2/N_\mathrm{dof}$ quantifying the quality of the data collapse for the two-point correlation function as a function of the critical exponents $\nu$ and $\eta$. We Taylor expand the scaling function of Eq.~\eqref{eq:fit_exponents} to $M=4$. The exact values are marked at the intersection of the two bold straight white lines. (c) Using the emulation data of (a). (d) Using the emulation data of (b).}
	\label{fig:fit_exponents}
\end{figure}

To evaluate the improvement on the estimated critical exponent values upon improving the hardware quality, we perform noisy emulations on intermediate-scale systems. In the main text, we determined that a noise parameter $p\approx 0.08$ reproduced the experimental data, see Figs.~2(e)--(f). Using the same form for the quantum circuit, we lower the noise level to $p=0.01$ and can obtain reliable data for $L=13$ and $6$ time steps using a first-order Suzuki-Trotter decomposition. Similarly, by lowering the noise further to $p=0.002$, we obtain data for $L=19$ and $12$ time steps. Leaving the exponents $\nu$ and $\eta$ as free parameters, we extract a region of maximum likelihood for their values in Figs.~\ref{fig:fit_exponents}(b)--(d). As expected, by reducing the noise level, we can access larger system sizes and longer drive times $T$, resulting in tighter bounds for the critical exponent values.

\section{Determining the noise-induced length scale}

\begin{figure}[h]
	\centering
	\includegraphics[width=0.4\textwidth]{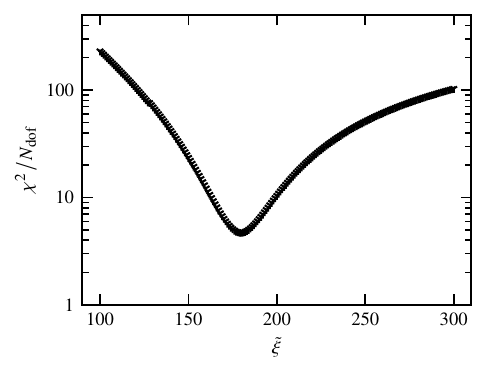}
	\caption{Chi-square per degree of freedom $\chi^2\bigl(\tilde{\xi}\bigr)/N_\mathrm{dof}$ for the data of Figs.~4(c)--(d) in the main text using the known values of the critical exponents, and leaving the noise-induced length scale $\tilde{\xi}$ as a free parameter in the fit. The best data collapse is obtained for $\tilde{\xi}\approx 180$.}
	\label{fig:fit_xi_noise}
\end{figure}

To determine the noise-induced length scale $\tilde{\xi}$ from Figs.~4(c)--(d) in the main text, we employ the same procedure used for determining the critical exponents. The scaling function of Eq.~\eqref{eq:fit_exponents} features a multiplicative term $\sim\exp(-xT/\tilde{\xi})$, with $\tilde{\xi}$ a free parameter in the fit. Here, we fix the critical exponents to their known values. The chi-square per degree of freedom is shown in Fig.~\ref{fig:fit_xi_noise} as a function of $\tilde{\xi}$, with a minimum found for $\tilde{\xi}\approx 180$. We use this value in the main text for correcting the data from Fig.~4(c) to Fig.~4(d).

\section{Additional data on the noise model}

\subsection{Noise dependence of the noise-induced length scale}

\begin{figure}[h]
	\centering
	\includegraphics[width=0.6\textwidth]{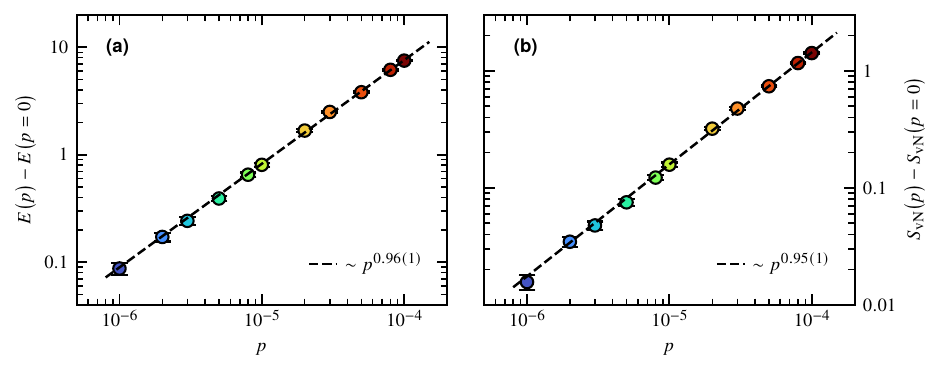}
	\caption{Emulation details: $L=33$ qubits with $36\,513$ gates comprised of Hadamard, $\hat{U}^x$, and $\hat{U}^{zz}$ with $T=32$, $t=0$, and second-order Suzuki-Trotter expansion where $\delta t=0.1$. The results are averaged over $\gtrsim 2\times 10^3$ random circuits. (a) Energy difference between finite $p$ and $p=0$ data as a function of $p$, showing a linear behavior $\propto p$. (b) Bipartite Von Neumann entanglement entropy difference between finite $p$ and $p=0$ data as a function of $p$, measured in the middle of the system $L/2$. It displays a linear behavior $\sim p$.}
	\label{fig:noise_energy_entropy}
\end{figure}

A microscopic picture for the noise model is that, in a given run, while the system is time-evolving, excitations in the form of Pauli operators $\hat{X}$, $\hat{Y}$, and $\hat{Z}$ are induced. To measure the excess of energy induced in the system, we define the energy,
\begin{equation}
    E=\langle\Psi(t)\vert\hat{\mathcal{H}}\bigl(T,t\bigr)\vert\Psi(t)\rangle.
    \label{eq:energy}
\end{equation}
Using matrix product states, we emulate quantum circuits for different noise strength $p$ with $L=33$ qubits and a fixed number of gates $36\,513$ gates comprised of Hadamard, $\hat{U}^x$, and $\hat{U}^{zz}$ with $T=32$, $t=0$, and second-order Suzuki-Trotter expansion where $\delta t=0.1$. The results are averaged over $\gtrsim 2\times 10^3$ random circuits. The excess of energy induced by the noise is the difference between $p$ and $p=0$ data, and is plotted in Fig.~\ref{fig:noise_energy_entropy}(a). We find that it grows linearly with $p$.

In addition, we measure the excess of bipartite Von Neumann entanglement entropy induced by the noise. The bipartite Von Neumann entanglement entropy between a subsystem $A$ comprised of the qubits $1,2,\ldots l$ and the rest of the system $B$ (qubits $l+1,l+2,\ldots L$) is defined as,
\begin{equation}
    S_\mathrm{vN}=-\mathrm{tr}_B\Bigl(\hat{\rho}_B\ln\hat{\rho}_B\Bigr),~~\mathrm{with}~\hat{\rho}_B=\mathrm{tr}_A\vert\Psi(t)\rangle\langle\Psi(t)\vert.
    \label{eq:entanglement_entropy}
\end{equation}
We measure the excess of bipartite Von Neumann entanglement entropy induced by the noise $p$ in the middle of the system at $L/2$, and plot it in Fig.~\ref{fig:noise_energy_entropy}(b). we find a linear scaling with $p$. This explains why it becomes increasingly difficult with the matrix product states, which rely on low entanglement, to emulate large quantum circuits as $p$ gets larger.

\subsection{Depth dependence of the noise-induced length scale}

\begin{figure}[h]
	\centering
	\includegraphics[width=0.6\textwidth]{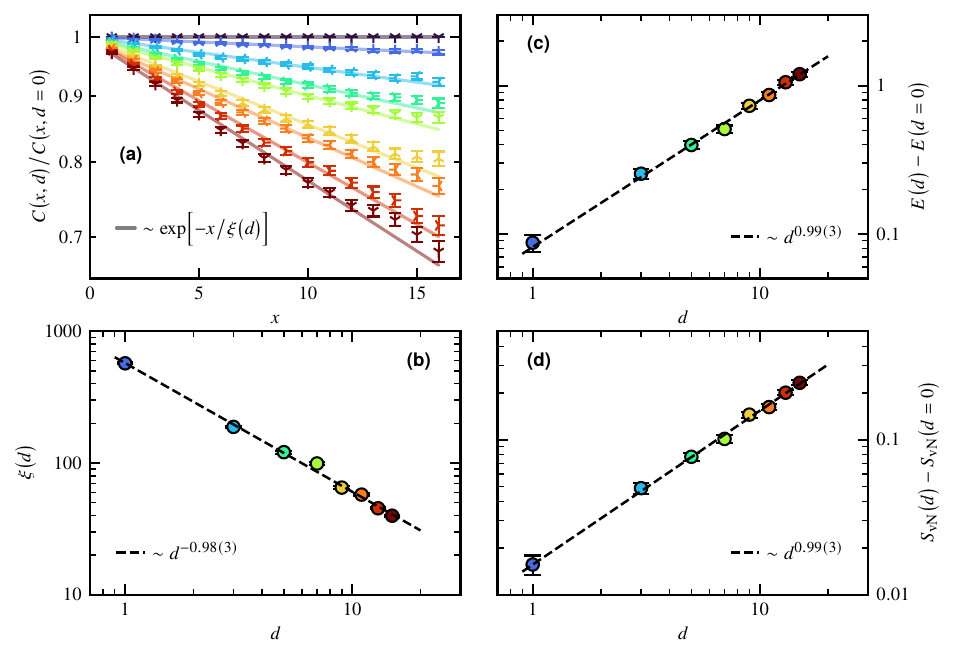}
	\caption{Emulation details: $L=33$ qubits, $p=10^{-6}$, $T=32$, $t=0$, and second-order Suzuki-Trotter expansion with $\delta t=0.1$. The $d=0$ data have $36\,513$ gates comprised of Hadamard, $\hat{U}^x$, and $\hat{U}^{zz}$ gates. The results are averaged over $\gtrsim 2\times 10^3$ random circuits. (a) Two-point correlation as a function of the distance $x$, rescaled by the $d=0$ data. Each curve corresponds to a value of the circuit depth $d$ whose code color can be read from the other panels. We fit the observed exponential decay $\sim\exp\bigl[-x/\xi(d)\bigr]$ (bold line), and extract the length $\xi(d)$. (b) Length $\xi(d)$ as a function of $d$, which shows a $\sim 1/d$ dependence. (c) Energy difference between finite $d$ and $d=0$ data as a function of $d$. (d) Bipartite Von Neumann entanglement entropy difference between finite $d$ and $d=0$ data as a function of $d$, measured in the middle of the system $L/2$. (c) and (d) both display a linear behavior with the depth $d$.}
	\label{fig:noise_depth_dependence}
\end{figure}

We established in the main text that at fixed circuit depth $d$, the noise induces a length scale $\xi(p)\sim 1/p$, with $p$ the strength of the noise. Here, we want to study at fixed $p$ the effect of the circuit depth $d$ on the noisy data. In practice, we increase the depth of the circuit by performing backward time steps in the evolution,
\begin{equation}
    \exp\bigl[-i\hat{\mathcal{H}}(T,t)\delta t\bigr] \to \Bigl(\exp\bigl[+i\hat{\mathcal{H}}(T,t)\delta t\bigr]\exp\bigl[-i\hat{\mathcal{H}}(T,t)\delta t\bigr]\Bigr)^{\frac{d-1}{2}}\exp\bigl[-i\hat{\mathcal{H}}(T,t)\delta t\bigr],
    \label{eq:eq:time_evo_depth}
\end{equation}
with $d=1$, $3$, $5$, etc. By definition, $d=0$ corresponds to the noiseless ($p=0$) data. The form of Eq.~\eqref{eq:eq:time_evo_depth} allows us to increase the circuit depth while keeping the same physics, making direct comparison of observables for different values of $d$ possible.

The form of the circuit is fixed with $L=33$ qubits and $p=10^{-6}$ ($T=32$, $t=0$, and second-order Suzuki-Trotter expansion with $\delta t=0.1$). The $d=0$ data have $36\,513$ gates comprised of Hadamard, $\hat{U}^x$, and $\hat{U}^{zz}$ gates. The results, plotted in Fig.~\ref{fig:noise_depth_dependence} are averaged over $\gtrsim 2\times 10^3$ random circuits. The first quantity we consider is the two-point correlation function as a function of the distance $x$, which we rescale by the reference $d=0$ data, as plotted in Fig.~\ref{fig:noise_depth_dependence}(a). We observe an exponential decay of the form $C\bigl(x,d\bigr)=C\bigl(x,d=0\bigr)\mathrm{e}^{-x/\xi(d)}$, with the depth-dependence of $\xi(d)$ shown in Fig.~\ref{fig:noise_depth_dependence}(b), where we find that $\xi(d)\sim 1/d$. It is compatible with the simple argument developed in the main text. Hence, both as a function of the noise strength $p$ and the depth $d$, we find that $\xi(p,d)\sim 1/pd$.

We also look at the excess of energy in Fig.~\ref{fig:noise_depth_dependence}(c), which grows linearly with $d$, and look at the excess of bipartite Von Neumann entanglement entropy, also scaling linearly with $d$, and plotted in Fig.~\ref{fig:noise_depth_dependence}(d). They display the same dependence as the circuit with fixed depth and varying noise strength $p$ considered above.

\subsection{Strong noise regime}

\begin{figure}[h]
	\centering
	\includegraphics[width=0.7\textwidth]{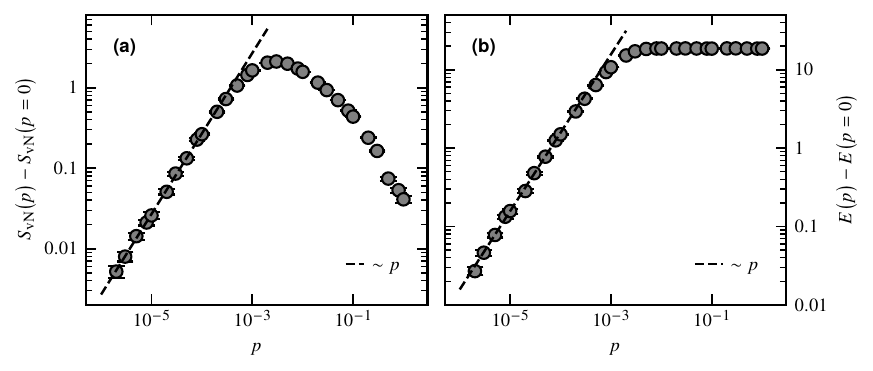}
	\caption{Emulation details: $L=15$ qubits, $T=14$, $t=0$, and second-order Suzuki-Trotter expansion with $\delta t=0.1$, corresponding to a total of $7\,155$ gates. (a) Bipartite Von Neumann entanglement entropy difference between finite $p$ and $p=0$ data as a function of $p$, measured in the middle of the system $L/2$. (b) Energy difference between finite $p$ and $p=0$ data as a function of $p$. They both display a linear behavior $\propto p$ for $p\lesssim 10^{-3}$.}
	\label{fig:strong_noise}
\end{figure}

The form of the circuit is fixed with $L=15$ qubits ($T=14$, $t=0$, and second-order Suzuki-Trotter expansion with $\delta t=0.1$), corresponding to a total of $7\,155$ gates. Through matrix product states simulations, we study the excess of bipartite Von Neumann entanglement entropy and energy induced by the noise up to $p=1$. The results are displayed in Fig.~\ref{fig:strong_noise}.

The low noise regime with the linear scaling $\propto p$ (previously studied) is observed for $p\lesssim p_\mathrm{c}\approx 2\times 10^{-3}$. Above this value, the two quantities of Figs.~\ref{fig:strong_noise}(a--b) display different behaviors. After reaching a maximum at $p_\mathrm{c}$, the excess of bipartite Von Neumann entanglement entropy decreases. The small $p$ regime can be understood as a heating regime with the noise analogous to induced excitations (see previous discussions). The decrease observed beyond the maximum as $p\to 1$ is associated with a slower growth for the entanglement resulting in an absolute smaller value for the entanglement in the fixed circuit considered.

As for the excess of energy, it shows a plateau as $E\bigl(p\gtrsim p_\mathrm{c}\bigr)\approx 0$. In absence of noise, the energy of the system is closed to its the ground state energy of order $-L$ (the energy is extensive). Seeing the noise-induced Pauli gates as excitations carrying a given amount of energy (order one constant), we find that $p_\mathrm{c}$ corresponds to a density of these excitations in the circuit of order one $\overline{N_\mathrm{e}(p_\mathrm{c})}/L\approx 1$, i.e., a value of noise strength such that on average each qubit has been affected by a noisy gate. With $\overline{N_\mathrm{e}}=pdL$ and $d$ the circuit depth, we find that $p_\mathrm{c}\approx 1/d$. For the data of Fig.~\ref{fig:strong_noise} where $d\approx 420$, the estimate $p_\mathrm{c}\approx 2\times 10^{-3}$ is compatible with the numerical observations.

In the limit of infinite system size and circuit depth, the stochastic noise model on top of the transverse field Ising model falls into the class of random unitary free fermion circuits~\cite{Dias2021} for $p>0$. Hence, we do not expect the two regimes of Fig.~\ref{fig:strong_noise}(a) to correspond---strictly speaking---to different phases. However, in a setup with a finite number of qubits and a finite circuit depth, one can observe different noise regimes as a function of $p$.

\section{Running on quantum hardware (Rigetti Aspen-9)}

\subsection{Hardware specificities}

\begin{figure}[h]
	\centering
	\includegraphics[width=0.7\textwidth]{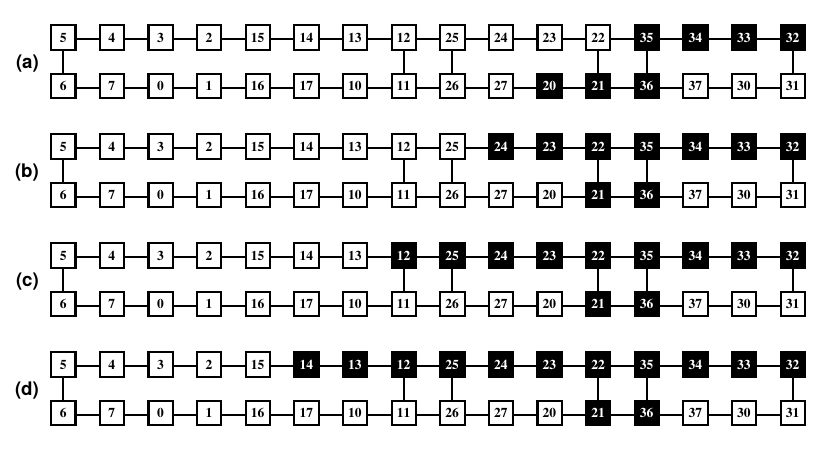}
	\caption{Lattice layout of the $32$ qubits Rigetti Aspen-9 chip. Each square represents a qubit with its index. The vertices represent the connectivity of the lattice, i.e., the pairs of qubits on which two-qubit gates can be applied. The qubits in black correspond to the ones that we use for the simulations, arranged in a continuous one-dimensional path: (a) $L=7$, (b) $L=9$, (c) $L=11$, and (d) $L=13$.}
	\label{fig:rigetti_aspen9_lattice}
\end{figure}

Rigetti quantum processors are based on tunable superconducting qubits. The Rigetti Aspen-9 chip features $32$ qubits with connectivity displayed in Fig.~\ref{fig:rigetti_aspen9_lattice}. The median $T_1=29.5~\mu$s and $T_2=17.5~\mu$s are to be compared with the gate duration multiplied by the circuit depth. A one-qubit gate duration is $\approx 50$~ns and a two-qubit gate duration $\approx 200$~ns. We evaluate that one can perform a few time steps before exceeding the coherence times of the qubits. We employ active reset of the qubits to the state $\vert 0\rangle$ after each run, which decreases the delay between running successive circuits (median fidelity of $99.5\%$).

Other performance numbers regarding the Rigetti Aspen-9 chip include the median one-qubit gate fidelity of $98.8\%$, the median two-qubit gate fidelity of $88.0\%$, and the median readout fidelity of $95.1\%$.

We model the noise with a depolarizing channel (see main text), which do not account for individual sources of errors. However, different noise models (depolarizing, bit-flip, phase-flip, and amplitude-damping channels) were studied in Appendix B of Ref.~\onlinecite{PRXQuantum.2.030346}. They were all found to induce an exponential decay when looking at a two-point correlation. This supports our choice of accounting for different sources of error through a model characterized by a single noise strength parameter $p$.

\subsection{Fidelity versus system size}

We have found that increasing the system size while maintaining the circuit depth constant lowers the quality of the data, as compared to the circuit emulations. Potential sources include, e.g., crosstalk, and the average quality of the qubits considered. We ran simulations on the quantum processor for $L=7$, $9$, $11$, and $13$ using the qubits highlighted in Fig.~\ref{fig:rigetti_aspen9_lattice}, and found that the data for $L=7$ provide the best agreement with the emulations. We document this observation in the following.

\begin{figure}[h]
	\centering
	\includegraphics[width=0.4\textwidth]{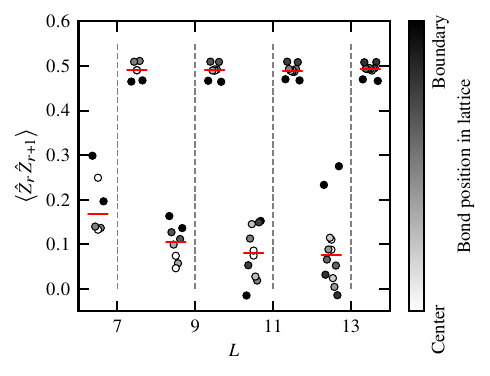}
	\caption{Two-point correlation function $\langle\hat{Z}_r\hat{Z}_{r+1}\rangle$ for all bonds $r=1,2,\ldots L-1$ for different system sizes $L$. Left of the dashed line corresponds to the quantum processor data. Right of the dashed line corresponds to the emulation. The different colors show the bond position in the system with respect to its center and boundary. The red horizontal line is the median of all the different bonds.}
	\label{fig:bond_correlation}
\end{figure}

We first consider the two-point correlation function $\langle\hat{Z}_r\hat{Z}_{r+1}\rangle$ for all bonds $r=1,2,\ldots L-1$ for different system sizes $L$. The drive time $T=1$ with time step $\delta t=0.5$ and a first-order Suzuki-Trotter decomposition is considered. The data are plotted in Fig.~\ref{fig:bond_correlation}. From the emulation, we expect the quantity to be roughly independent of the system size $L$ with a median value $\approx 0.5$. In the experimental data, the problem is not so much that the data have a lower median value (this is, e.g., captured by the noise model), but that the median value decreases as the system size increases. The two-point correlation function versus the distance $x$ between two qubits is a decreasing function of $x$: hence if the $x=1$ data are already relatively small in amplitude, there is little chance to capture the correlation for $x>1$ as it lies already too close to zero, beyond the resolution controlled by the number of outputs when computing the observable. This is why we focused on the $L=7$ simulation data for the analysis.

\begin{figure}[t]
	\centering
	\includegraphics[width=0.6\textwidth]{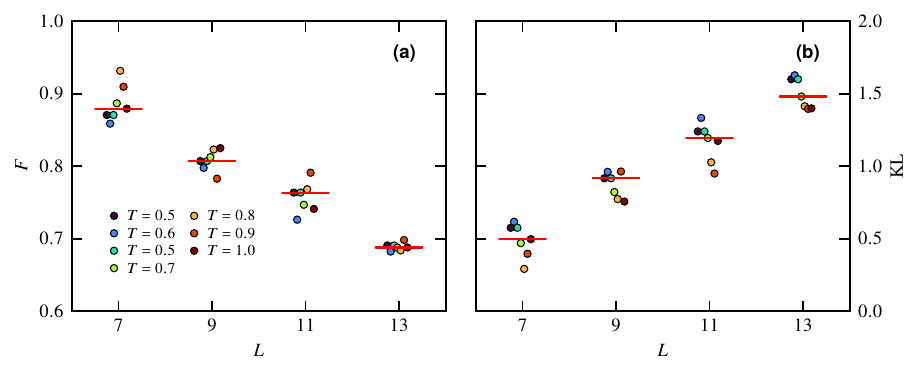}
	\caption{The data points correspond to different values of the drive time $T$ with time step $\delta t=T/2$ and a first-order Suzuki-Trotter decomposition. The horizontal line corresponds to the median. (a) Fidelity as defined in Eq.~\eqref{eq:fidelity} as a function of the system size. (b) Kullback–Leibler divergence as defined in Eq.~\eqref{eq:kl_divergence} as a function of the system size.}
	\label{fig:fidelity_kl}
\end{figure}

We also consider two others quantities to evaluate the quality of the experimental data as a function of the system size $L$. The first one is the fidelity which we define as,
\begin{equation}
    F = \sum\nolimits_{\{\boldsymbol{\sigma}\}}\sqrt{p_{\boldsymbol{\sigma}}^\mathrm{exact}\times p_{\boldsymbol{\sigma}}^\mathrm{sampled}},
    \label{eq:fidelity}
\end{equation}
with $\vert\Psi\rangle=\sum\nolimits_{\{\boldsymbol{\sigma}\}}c_{\boldsymbol{\sigma}}\vert\boldsymbol{\sigma}\rangle$, and $p_{\boldsymbol{\sigma}}=\vert{c}_{\boldsymbol{\sigma}}\vert^2$ where $\sum\nolimits_{\{\boldsymbol{\sigma}\}}p_{\boldsymbol{\sigma}}=1$. The sum runs over the $2^L$ basis states $\{\boldsymbol{\sigma}\}$. Note that $p_{\boldsymbol{\sigma}}^\mathrm{sampled}$ is constructed from a finite number of samples by counting the number of bitstrings $\boldsymbol{\sigma}$ and normalizing by the total number of samples. On the other hand, $p_{\boldsymbol{\sigma}}^\mathrm{exact}$ can be obtained exactly by emulating the quantum circuit without need to sample. The second quantity that we  define is the Kullback–Leibler (KL) divergence,
\begin{equation}
    \mathrm{KL} = \sum\nolimits_{\{\boldsymbol{\sigma}\}}p_{\boldsymbol{\sigma}}^\mathrm{sampled}\ln\left(\frac{p_{\boldsymbol{\sigma}}^\mathrm{sampled}}{p_{\boldsymbol{\sigma}}^\mathrm{exact}}\right),
    \label{eq:kl_divergence}
\end{equation}
which measures of how the sampled distribution is different from the exact one.

The results are plotted in Fig.~\ref{fig:fidelity_kl}. Because both quantities are global, we expect that the effect of noise will increase as the system size increases. It is confirmed by the decreasing fidelity and the increasing KL divergence, and supports the fact that $L>7$ simulation data are of lower quality.

\end{document}